\documentclass[a4paper,11pt]{article}
\usepackage{pos}

\usepackage[shortcuts]{extdash}
\newcommand\LST{LST\=/1}
\usepackage{gensymb}
\usepackage{physics}

\title{Performance of the Large-Sized Telescope prototype of the Cherenkov Telescope Array}
 \ShortTitle{Performance of the CTA LST-1}

\author*[a,b]{D. Morcuende}
\author[b]{R. López-Coto}
\author[c]{A. Moralejo}
\author[d]{S. Nozaki}
\author[e]{T. Vuillaume}

\affiliation[a]{IPARCOS-UCM, Instituto de Física de Partículas y del Cosmos, and EMFTEL Department, Universidad Complutense de Madrid,\\
E-28040 Madrid, Spain}

\affiliation[b]{Instituto de Astrofísica de Andalucía-CSIC,\\
Glorieta de la Astronomía s/n, 18008, Granada, Spain}

\affiliation[c]{Institut de Fisica d'Altes Energies (IFAE), The Barcelona Institute of Science and Technology,\\
Campus UAB, 08193 Bellaterra (Barcelona), Spain}

\affiliation[d]{Max-Planck-Institut für Physik,\\
Föhringer Ring 6, 80805 München, Germany}

\affiliation[e]{Univ. Savoie Mont Blanc, CNRS, Laboratoire d'Annecy de Physique des Particules - IN2P3,\\
74000 Annecy, France}

\onbehalf{on behalf of the CTA-LST Project} 

\emailAdd{dmorcuende@iaa.es}

\abstract{
The next-generation ground-based gamma-ray Cherenkov Telescope Array Observatory (CTAO) will consist of imaging atmospheric Cherenkov telescopes (IACTs) of three different sizes distributed in two sites. The Large-Sized Telescopes will cover the low-energy end of the CTA energy range, starting at about 20 GeV. After its first years of operation at the CTA northern site, the Large-Sized Telescope prototype (\LST{}) is in the final stage of its commissioning phase, having collected a significant amount of scientific data to date.

In this contribution, we present the physics performance of the telescope using low-zenith Crab Nebula observations and Monte Carlo simulations fine-tuned accordingly. We show performance figures of merit such as the energy threshold, effective area, energy and angular resolution, and sensitivity based on the standard Hillas-parameters approach and following the source-independent and dependent analysis methods. The analysis threshold is estimated at 30 GeV. The energy resolution is around 30\%, and the angular resolution is 0.3 degrees at 100 GeV. The best integral sensitivity of \LST{} is about 1.1\% of the Crab Nebula flux above 250 GeV for 50 hours of observations. We also show the spectral energy distribution and light curve from Crab Nebula observations, which agree with results from other IACTs and link smoothly with \textit{Fermi}-LAT when considering statistical and systematic uncertainties near the energy threshold.
}

\ConferenceLogo{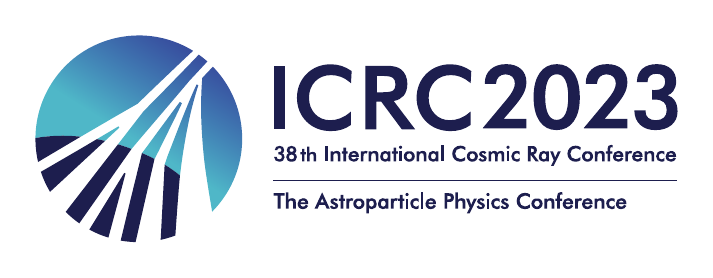}

\FullConference{%
38th International Cosmic Ray Conference (ICRC2023)\\
  26 July - 3 August, 2023\\
  Nagoya, Japan}

\begin{document}
\maketitle

\section{Introduction}

The Large-Sized Telescopes (LSTs) are the largest among the imaging air Cherenkov telescopes (IACTs) that will comprise the Cherenkov Telescope Array Observatory (CTAO), a next-generation ground-based facility for VHE gamma-ray astronomy. LSTs are designed to cover the lower end of the energy range targeted by CTA, down to energies of about 20 GeV. The {LST prototype}, named \LST{}, has been operational {at} the North site of CTAO, at the Observatorio Roque de Los Muchachos (La Palma, Spain), since November 2019, and has been frequently observing the Crab Nebula during its commissioning phase, among other astrophysical sources. VHE gamma-ray emission from the Crab Nebula has been thoroughly characterized over the past decades and found to be stable within the systematic uncertainties of the VHE instruments, hence it is considered a calibration source.

In this contribution, {we characterize the \LST{} performance based on Crab Nebula observations and validate} the Monte Carlo (MC) simulations used for the \LST{} data analysis. We compare the results obtained on the Crab Nebula spectrum and light curve with those reported previously by other instruments. Furthermore, we calculate the differential sensitivity of the \LST{} operating as a single telescope. For a more detailed description of this work, see \cite{LST1_performance}.

\section{Monte Carlo simulations}
Detailed MC simulations are key in the reconstruction of the air shower properties (type of particle, energy and direction) and to produce the instrument response functions (IRFs) for high-level analysis. We used \texttt{CORSIKA} v7.7100 \cite{corsika} to simulate shower development and Cherenkov light emission, and \texttt{sim\_telarray} v2020-06-28 \cite{simtelarray} for telescope simulation, which includes the reflection of Cherenkov light on the mirror dish, its passage through the camera entrance window, and its recording by the camera. These simulations are analyzed using the same pipeline as the observed data.

A first set of simulations consisting of gamma rays and protons is used to train {random forest (RF) models \cite{breiman}} {for the reconstruction of the physics properties of the observed gamma rays}. These models are also applied to another MC sample of gamma rays (so-called test sample) to calculate the IRFs to be used in the high-level analysis of the selected gamma-ray events. Both MC training and test samples were produced in a grid of different telescope-pointing directions following the path of the Crab Nebula as observed by \LST{}. Subsequently, we added the pointing information to the set of training parameters of the RF models so that it was considered in the reconstruction. Furthermore, the IRFs used for the analysis of each data run correspond to its nearest pointing node in the MC grid.

We adjusted the MC simulation to match the real optical efficiency (which shows a $\pm5\%$ variation in the time interval of the observations), monitored using muon ring images, and the optical point spread function of the telescope. {The analysis presented here does not use a run-wise tuning, but MC tuned to the average observational conditions of the data set.} Also, we adjusted the night sky background level in the MC simulation to the one found in the Crab Nebula field of view (FoV). {The comparison of the distributions of the image parameters for both MC simulations and observed gamma-ray excess from the Crab Nebula yields an excellent agreement.}

\subsection{Instrument response functions}
\begin{figure}
    \centering
    \includegraphics[width=0.49\linewidth]{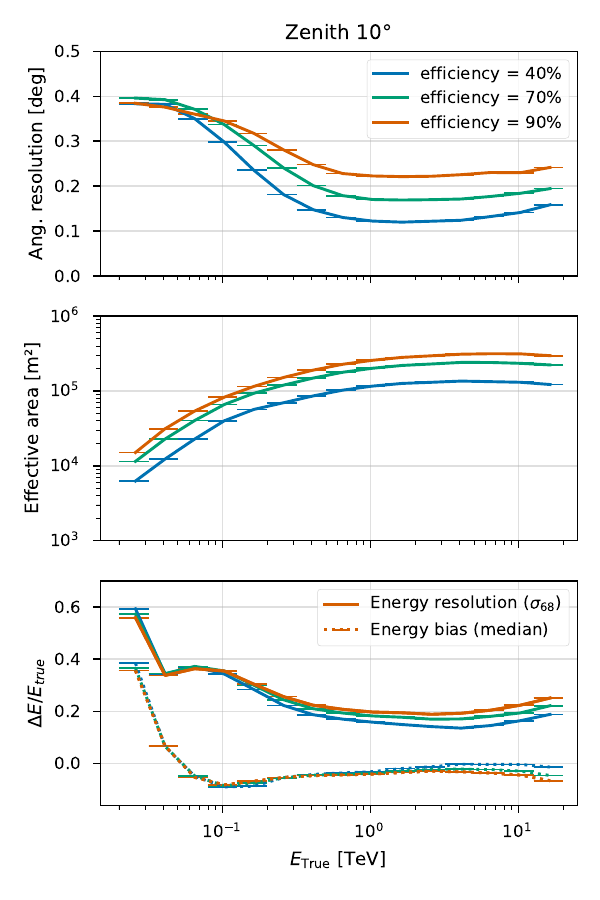}
    \includegraphics[width=0.49\linewidth]{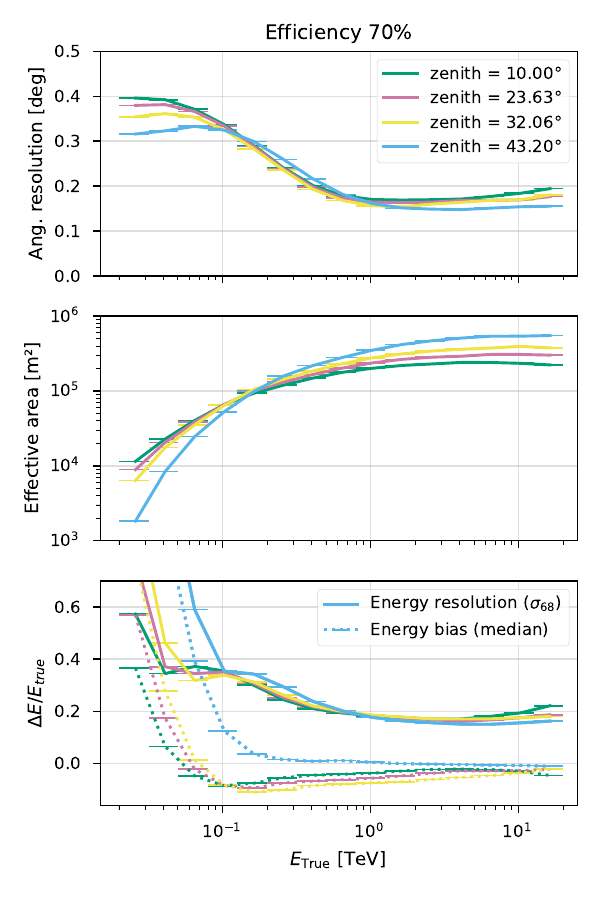}
    \vspace{-0.3cm}
    \caption{IRFs as a function of the true energy for source-independent analysis. Left panels: fixed zenith angle at $10\degree$ and several gamma-ray selection efficiencies. Right panels: fixed gamma-ray selection efficiency to 70\% for several zenith angles. Top panels: the angular resolution, mid panels: the effective area, and bottom panels: the energy resolution and bias. The angular cut for the calculation of the effective area and energy resolution has an efficiency of 70\%. Figure from \cite{LST1_performance}.
    }
    \label{fig:irfs_srcindep}
\end{figure}

IRFs (effective collection area, angular and energy resolution) are essential figures of merit in the telescope's performance. {We calculated these quantities based on the test MC sample as a function of the true energy. Figure~\ref{fig:irfs_srcindep} shows the dependence for a fixed zenith angle of $10\degree$ and varying efficiencies of gamma-ray selection cuts, and vice versa for a fixed efficiency of 70\% and varying zenith angles.}

\section{Crab data sample and analysis pipeline}

We used {34.2 hours of effective observation time} of good-quality Crab Nebula data, {consisting of} observations carried out at low zenith angles (below $35\degree$), in dark night conditions, and in good weather, in order to achieve the lowest energy threshold possible. The data were recorded between November 2020 and March 2022, spanning $\simeq1.5$~years. The observations were performed in {wobble mode \cite{FOMIN1994}, pointing in two mirrored directions $0.4\degree$ away from the location of the Crab Nebula.}

The low-level processing of the \LST{} data from raw uncalibrated waveforms up to lists of gamma-ray-like events (DL3) was done with the analysis library \texttt{cta-lstchain} v0.9 \cite{lst_performance_icrc2021}, {built based on} \texttt{ctapipe} \cite{ctapipe_ICRC_2023}. We calculated the IRFs using \texttt{pyirf} \cite{pyirf_v0.7}. The handling of the reduction of the observed data was performed using \texttt{lstosa} \cite{lstosa_ADASS}, while the training of models and production of the IRFs was managed through \texttt{lstMCpipe} \cite{garcia2022lstmcpipe}. We followed the classic source-independent Hillas-based analysis and RF event-reconstruction techniques. The monoscopic analysis is usually limited by the background rejection, especially near the energy threshold. To enhance the signal reconstruction, at these energies, besides the standard source-independent analysis, we also use the source-dependent approach, which takes advantage of the a priori knowledge of the source location.

\section{Results}
In this section, we {present} the spectral energy distribution (SED) and light curve of the Crab Nebula, comparing them with measurements by other instruments as an alternative way to evaluate the performance of the telescope. Finally, we {show} the differential flux sensitivity of \LST{} for point-like source observations also using the Crab Nebula observations.

\subsection{Crab Nebula spectrum and light curve}
We used Gammapy v0.20 \cite{Gammapy_proceeding} to compute the SED and light curve. We first select events with image intensity greater than 80 photoelectrons, a cut that ensures a good agreement between real data and simulations by removing events too close to the trigger threshold, which was not completely stable during the commissioning phase. Then, we select gamma rays by applying energy-dependent gammaness {and directional cuts (on the so-called $\theta$ and $\alpha$ parameters for the source-independent and source-dependent analyses, respectively)} with a given efficiency, which leaves a certain percentage of gamma-ray MC events in each bin. We use 70\% efficiency in both gammaness and angular cuts for the baseline analysis.

\begin{figure}
    \centering
    \includegraphics[width=0.49\textwidth]{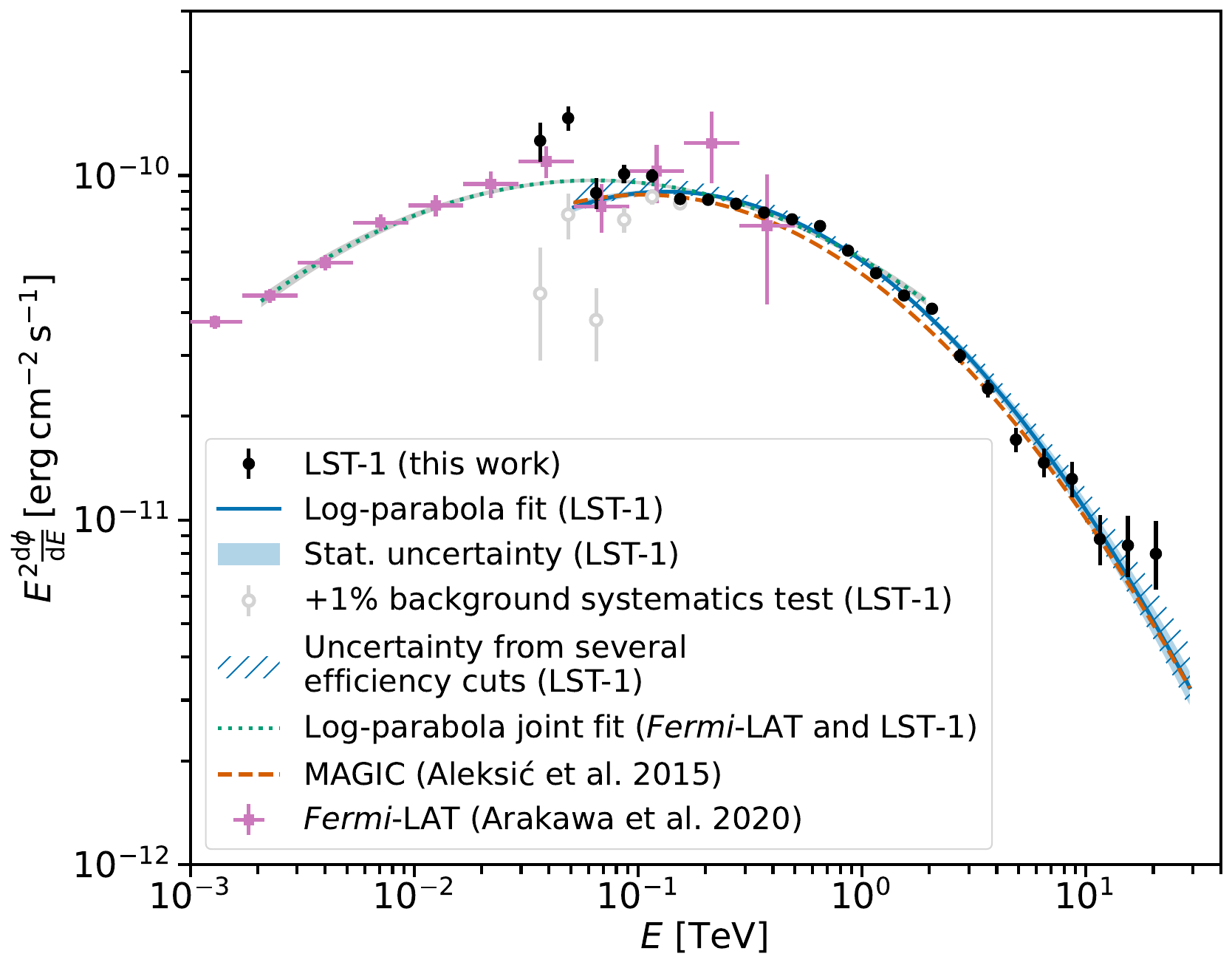}
    \includegraphics[width=0.49\textwidth]{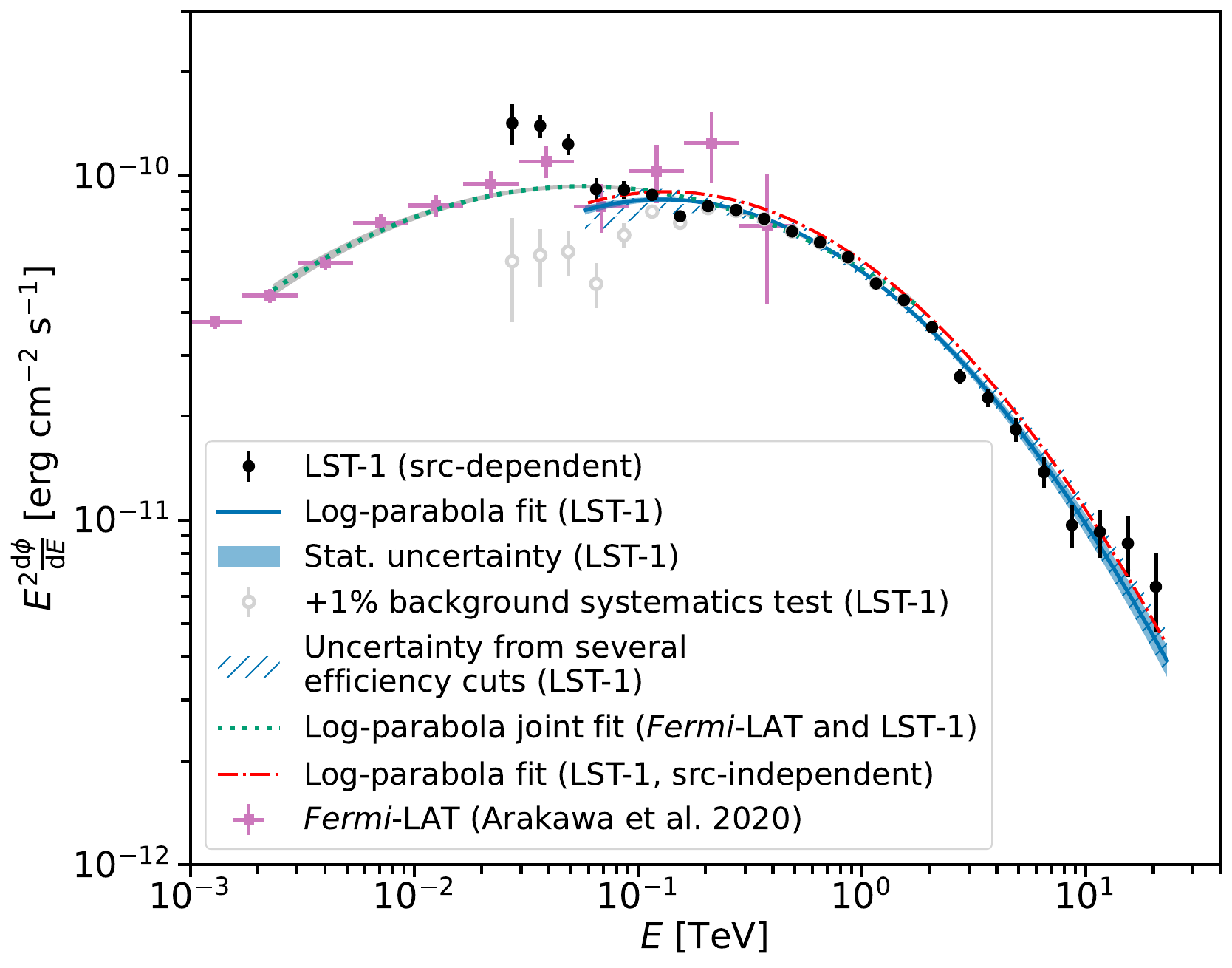}
    \vspace{-0.3cm}
    \caption{SED of the Crab Nebula obtained using the source-independent (left) and source-dependent (right) analysis. Flux points (black circles) and best-fit model (solid line) correspond to the dataset with a cut in image intensity > 80 p.e., and energy-dependent \textit{gammaness} and $\alpha/\theta$ selection cuts with 70\% gamma-ray efficiency. Figure from \cite{LST1_performance}.}
    \label{fig:crab_spectrum}
\end{figure}

To extract the gamma-ray signal and estimate the residual background, we use aperture photometry with a single control off-source sky region within the same FoV as the on-source region from which the signal is extracted. We calculate the spectrum through a forward-folding likelihood fit from 50~GeV to 30~TeV assuming a log-parabolic model:
$
    \dv*{\phi}{E}=f_{0} \cdot (E/E_{0})^{-\alpha-\beta \cdot \ln(E/E_{0})} \,[{\rm cm}^{-2}\\ \,{\rm s}^{-1} \,{\rm TeV}^{-1}] \rm{,}
$
with $E_0=400$~GeV, close to the decorrelation energy. The resulting best-fit model for {the} source-independent and source-dependent analyses is shown in Fig.~\ref{fig:crab_spectrum} and the {determined} spectral parameters are listed in Table~\ref{tab:spectral_parameters}. Based on these spectral models, flux points {are computed} by fitting only the normalization in each energy bin. {To test the consistency of the obtained Crab Nebula SED near the \LST{} threshold with that from \textit{Fermi}-LAT, we performed a joint fit from 2~GeV to 2~TeV, using \LST{} data and \textit{Fermi}-LAT spectral points \cite{Arakawa_2020}.} {\LST{} spectral} points below 50~GeV are computed based on this joint-model fit.

\begin{table}
    \centering
    \begin{tabular}{c|c|c|c|c}
Analysis type & $f_0$ [TeV$^{-1}$cm$^{-2}$ s$^{-1}$] & $\alpha$ & $\beta$ & $\chi^{2}/N_{\rm dof}$ \\
\hline
\hline
Source-independent & $(3.05 \pm 0.02) \times 10^{-10}$ & $2.25 \pm 0.01$ & $0.114 \pm 0.006 $ & 48.5 / 18\\ 
Source-dependent   & $(2.87 \pm 0.02) \times 10^{-10}$ & $2.26 \pm 0.01$ & $0.115 \pm 0.006 $ & 32.9 / 18\\ 
    \end{tabular}
    \vspace{-0.3cm}
    \caption{
    Spectral parameters of the \LST{} Crab Nebula best-fit model in the energy range 50 GeV - 30 TeV assuming a log-parabolic model. In both cases $E_0=400$~GeV and quoted uncertainties are only statistical. 
    }
    \label{tab:spectral_parameters}
\end{table}

We tested the stability of the resulting SED against {the choice of the cut values by assuming a grid of gamma-ray selection} cut efficiencies of 40\%-90\% for gammaness and 70\%-90\% for $\theta/\alpha$ cuts. We illustrate the band containing all the resulting SEDs fits assuming different combinations of the efficiencies in Fig.~\ref{fig:crab_spectrum}. Additionally, given the apparent discrepancy of the first spectral points of the \LST{} from the best-fit model, assuming only statistical uncertainties, we checked the effect of a possible systematic uncertainty in the estimation of the background. We tested the effect {of a $+1\%$ change in} the background normalization. This slight modification has a great impact on the low-energy spectral points, as shown by the open markers in Fig.~\ref{fig:crab_spectrum}. Actually, a change of $+0.5\%$ in the background normalization is sufficient to make the first \LST{} points compatible with the joint \textit{Fermi}-LAT+LST-1 SED in the case of the Crab Nebula. This agrees with the fact that we see differences of $\approx0.5\%$ in the background rate between different equidistant off-source positions.

{Considering these uncertainties,} the \LST{} spectrum agrees with that reported by MAGIC in the same energy range \cite{CrabMAGIC} within 10\% in flux, which is typically the level of systematic uncertainty {of IACTs}. Moreover, the \LST{} spectrum {at low energies agrees with} that reported by \textit{Fermi}-LAT.

\begin{figure}
    \centering
    \includegraphics[width=0.49\textwidth]{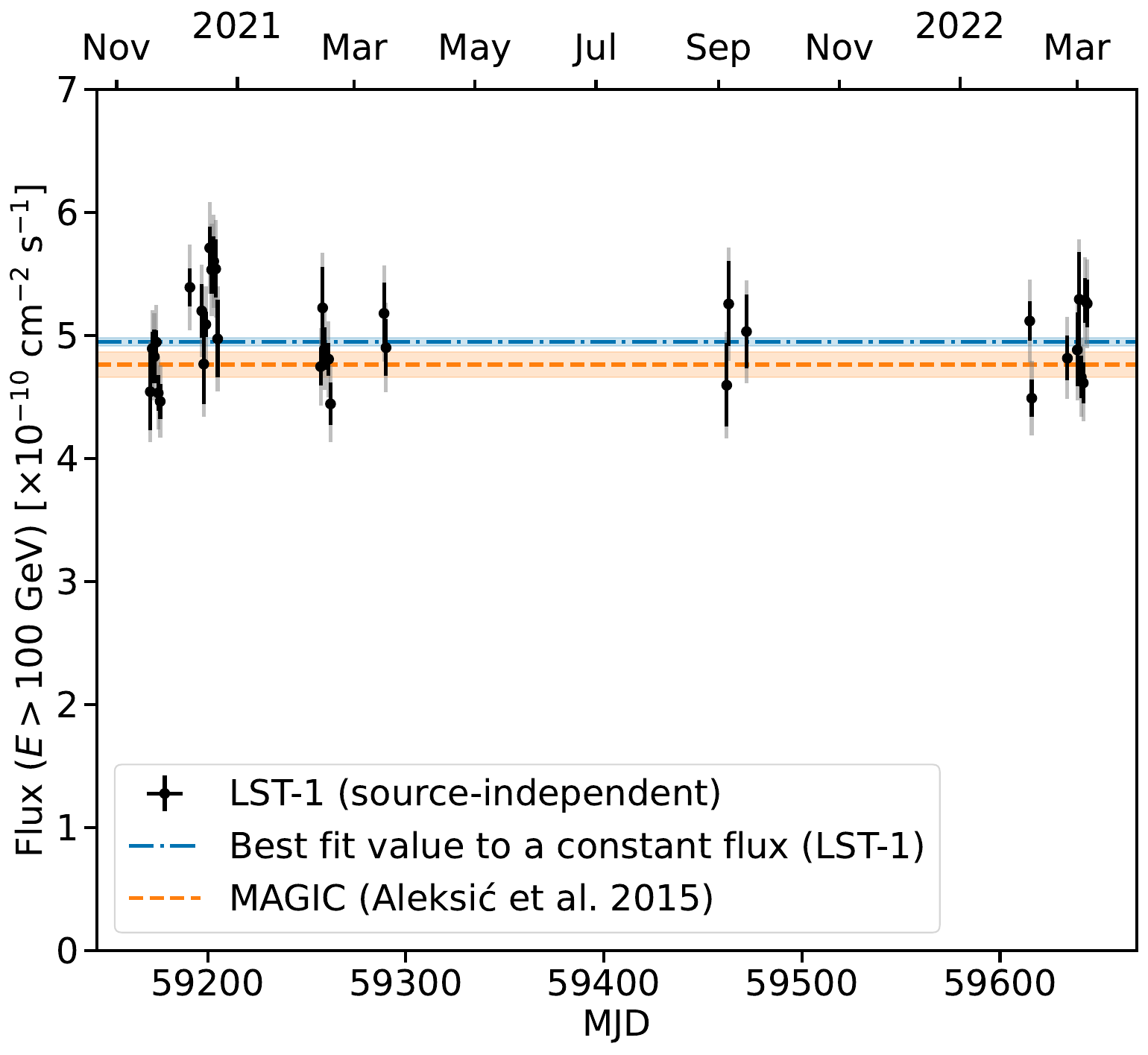}
    \includegraphics[width=0.49\textwidth]{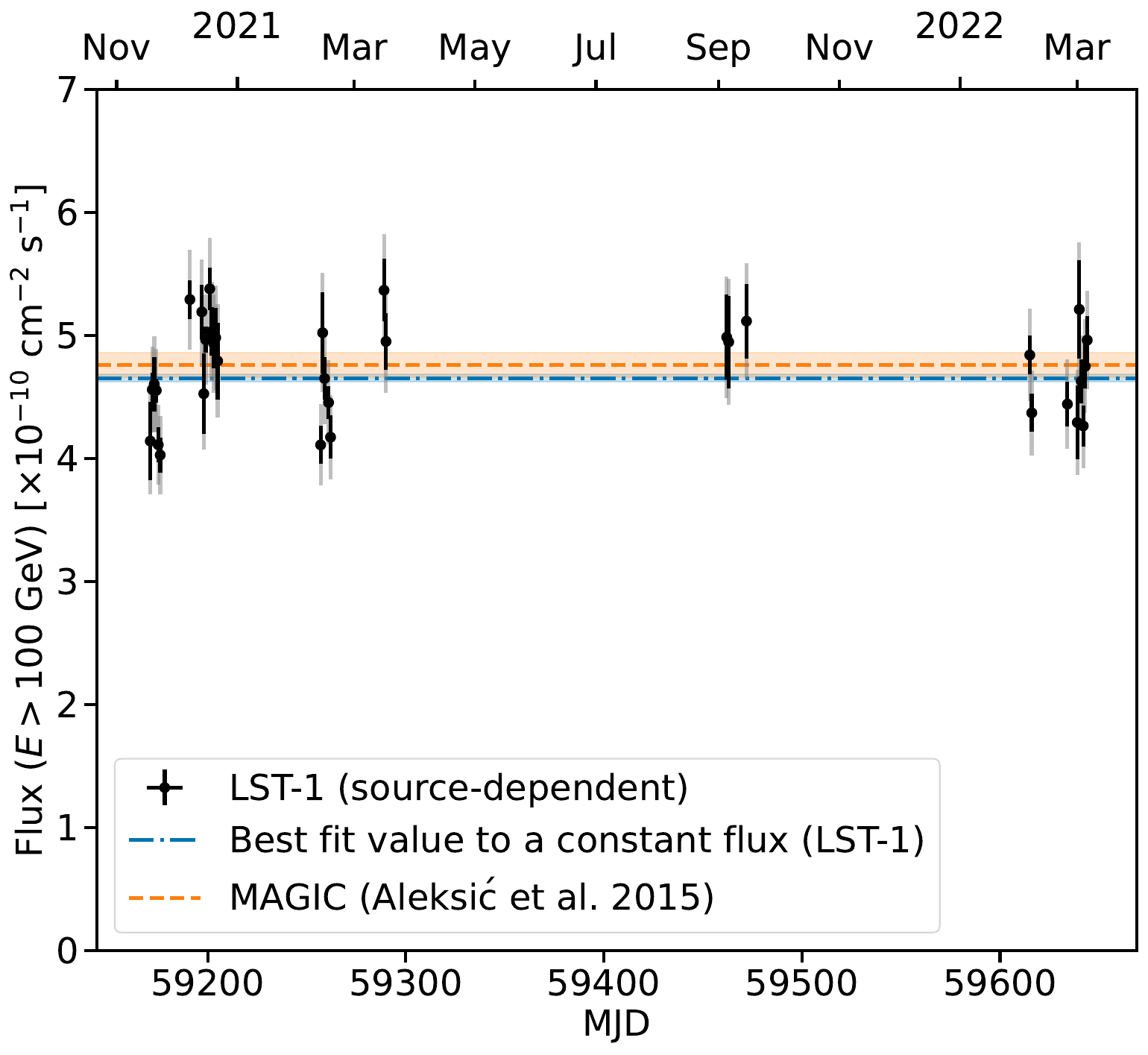}
    \vspace{-0.3cm}
    \caption{Light curve (>100~GeV) of the Crab Nebula using 1-day bins for source-independent (left) and source-dependent analyses (right). The best fit to a constant flux is plotted as a dash-dotted line. For reference, we also depict the integral flux in the same energy range calculated from \cite{CrabMAGIC}. The black error bars correspond to the statistical errors, while the gray ones include the systematic uncertainties summed quadratically (see main text). Figure taken from \cite{LST1_performance}.}
    \label{fig:crab_lc}
\end{figure}

To check that the VHE gamma-ray flux from the Crab Nebula remained stable across the time span of observations used in this work, we computed the daily light curve above 100~GeV (Fig.~\ref{fig:crab_lc}). We assume a log-parabola spectral model with the best-fit spectral parameters reported in Table~\ref{tab:spectral_parameters}. We test the steady flux hypothesis by fitting the light curve to a constant value of $F_{>100 \rm{~GeV}} = (4.95 \pm 0.03) \times 10^{-10} \rm{~cm}^{-2} \rm{~s}^{-1}$, with $\chi^2 / N_{dof} = 119.2 / 33$ (P-value=$1 \times 10^{-11}$) for the source-independent analysis, and $F_{>100 \rm{~GeV}} = (4.65 \pm 0.03) \times 10^{-10} \rm{~cm}^{-2} \rm{~s}^{-1}$, with $\chi^2 / N_{dof} = 147.6 / 33$ (P-value=$2 \times 10^{-16}$) for the source-dependent analysis. Only assuming statistical uncertainties, the results are clearly {inconsistent} with constant flux. Nonetheless, the tests conducted on the SED reveal that the total uncertainty must be significantly higher. We calculated that the additional uncertainty (added in quadrature to the statistical uncertainty) to make the light curves compatible with a stable flux is 6\% and 7\% on the night-wise flux values, respectively, for the source-independent and source-dependent analyses. This level of night-to-night systematics is not surprising, as we did not account for possible variable observing conditions in this work by using run-wise or night-wise~IRFs.

\subsection{Flux sensitivity}
Flux sensitivity is the minimum flux from a point-like source that {an IACT} can detect with a statistical significance of 5\,$\sigma$ assuming 50-hour observation time, and using 5~OFF regions for the background estimation. Moreover, we impose a minimum of 10 gamma rays detected in each energy bin (we use 5 logarithmic bins per decade) and a signal-to-background ratio ($S/B$) larger than 5\%. The sensitivity calculation is {based on the low-zenith Crab Nebula data, and follows} the source-independent and source-dependent methods. Angular and gammaness selection cuts were optimized on a subset of the Crab Nebula data to achieve the best sensitivity in each bin of energy.

\begin{figure}
    \centering
    \includegraphics[width=0.6\textwidth]{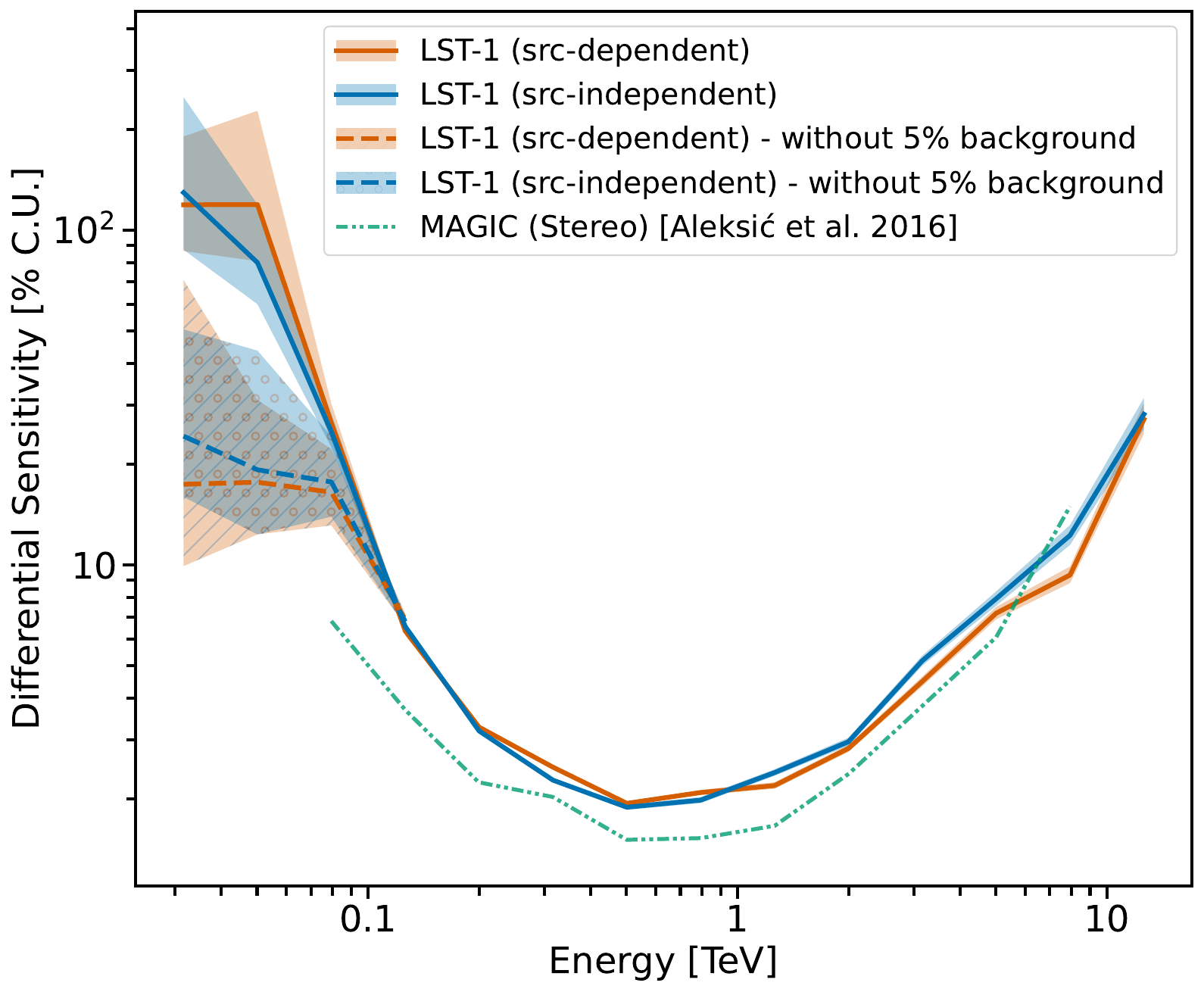}
    \vspace{-0.3cm}
    \caption{Differential sensitivity for source dependent and source independent analyses, as a function of reconstructed energy, with and without including the condition that the signal-to-noise ratio has to be at least 5\%. The MAGIC reference is taken from \cite{ALEKSIC201676}. Figure from \cite{LST1_performance}.}
    \label{fig:diff_sensitivity}
\end{figure}

Figure \ref{fig:diff_sensitivity} depicts the resulting sensitivity, with bands indicating the total uncertainty, assuming a 1\% systematic uncertainty in background normalization. We additionally show the sensitivity in flux without considering the requirement {to have a $S/B>5\%$}. This indicates {that the sensitivity near the threshold is limited by the systematic uncertainty of the background normalization}. The two analysis methods considered in this work result in similar flux sensitivity, with the source-dependent approach providing marginally better results above 1~TeV.

The optimal integrated sensitivity in 50 hours of observation is achieved for energies greater than 250 GeV, yielding 1.1\% Crab units. Also, since one of the main objectives of the \LST{} is detecting transients in short observing times, we estimate the integrated sensitivity for an exposure of 0.5 hours above 250 GeV, which is 12.4\% Crab units. In comparison, we show the sensitivity curve of MAGIC, which on average is a factor of 1.5 better above 100 GeV. This difference is due to the benefit of the stereo reconstruction and distant muon rejection despite the lower energy threshold of \LST{}.

\section{Conclusions}

{We report the performance of the \LST{} estimated based on Crab Nebula observations during its commissioning} and the MC simulations used for the analysis. The detailed description of the study can be found in~\cite{LST1_performance}. While we are able to properly reproduce the Crab Nebula spectrum and light curve, we remark the importance of understanding the systematic uncertainties near the threshold. 
A separate contribution focuses on the Crab pulsar analysis~\cite{LST1_pulsars_ICRC2023}, emphasizing the exceptional capabilities of \LST{} near its energy threshold at a few tens of GeV. 

Until the upcoming {CTAO-North comes} into play in the next couple of years, \LST{} is already a competitive instrument able to produce scientific results (see, e.g.,~\cite{LHAASO_LST1}). \LST{} is also exploiting the possibility of jointly observing with the MAGIC telescopes located nearby. Simultaneous observations allow for a stereoscopic system of three telescopes with higher sensitivity than MAGIC and \LST{} {alone} \cite{MAGIC_LST1_performance_ICRC23}.

\footnotesize{
\bibliographystyle{JHEP}
\bibliography{bibliography.bib}
}




\section*{Acknowledgements}
\tiny
We gratefully acknowledge financial support from the following agencies and organisations:

\bigskip
Conselho Nacional de Desenvolvimento Cient\'{\i}fico e Tecnol\'{o}gico (CNPq), Funda\c{c}\~{a}o de Amparo \`{a} Pesquisa do Estado do Rio de Janeiro (FAPERJ), Funda\c{c}\~{a}o de Amparo \`{a} Pesquisa do Estado de S\~{a}o Paulo (FAPESP), Funda\c{c}\~{a}o de Apoio \`{a} Ci\^encia, Tecnologia e Inova\c{c}\~{a}o do Paran\'a - Funda\c{c}\~{a}o Arauc\'aria, Ministry of Science, Technology, Innovations and Communications (MCTIC), Brasil;
Ministry of Education and Science, National RI Roadmap Project DO1-153/28.08.2018, Bulgaria;
Croatian Science Foundation, Rudjer Boskovic Institute, University of Osijek, University of Rijeka, University of Split, Faculty of Electrical Engineering, Mechanical Engineering and Naval Architecture, University of Zagreb, Faculty of Electrical Engineering and Computing, Croatia;
Ministry of Education, Youth and Sports, MEYS  LM2015046, LM2018105, LTT17006, EU/MEYS CZ.02.1.01/0.0/0.0/16\_013/0001403, CZ.02.1.01/0.0/0.0/18\_046/0016007 and CZ.02.1.01/0.0/0.0/16\_019/0000754, Czech Republic; 
CNRS-IN2P3, the French Programme d’investissements d’avenir and the Enigmass Labex, 
This work has been done thanks to the facilities offered by the Univ. Savoie Mont Blanc - CNRS/IN2P3 MUST computing center, France;
Max Planck Society, German Bundesministerium f{\"u}r Bildung und Forschung (Verbundforschung / ErUM), Deutsche Forschungsgemeinschaft (SFBs 876 and 1491), Germany;
Istituto Nazionale di Astrofisica (INAF), Istituto Nazionale di Fisica Nucleare (INFN), Italian Ministry for University and Research (MUR);
ICRR, University of Tokyo, JSPS, MEXT, Japan;
JST SPRING - JPMJSP2108;
Narodowe Centrum Nauki, grant number 2019/34/E/ST9/00224, Poland;
The Spanish groups acknowledge the Spanish Ministry of Science and Innovation and the Spanish Research State Agency (AEI) through the government budget lines PGE2021/28.06.000X.411.01, PGE2022/28.06.000X.411.01 and PGE2022/28.06.000X.711.04, and grants PID2022-139117NB-C44, PID2019-104114RB-C31,  PID2019-107847RB-C44, PID2019-104114RB-C32, PID2019-105510GB-C31, PID2019-104114RB-C33, PID2019-107847RB-C41, PID2019-107847RB-C43, PID2019-107847RB-C42, PID2019-107988GB-C22, PID2021-124581OB-I00, PID2021-125331NB-I00; the ``Centro de Excelencia Severo Ochoa" program through grants no. CEX2019-000920-S, CEX2020-001007-S, CEX2021-001131-S; the ``Unidad de Excelencia Mar\'ia de Maeztu" program through grants no. CEX2019-000918-M, CEX2020-001058-M; the ``Ram\'on y Cajal" program through grants RYC2021-032552-I, RYC2021-032991-I, RYC2020-028639-I and RYC-2017-22665; the ``Juan de la Cierva-Incorporaci\'on" program through grants no. IJC2018-037195-I, IJC2019-040315-I. They also acknowledge the ``Atracción de Talento" program of Comunidad de Madrid through grant no. 2019-T2/TIC-12900; the project ``Tecnologi\'as avanzadas para la exploracio\'n del universo y sus componentes" (PR47/21 TAU), funded by Comunidad de Madrid, by the Recovery, Transformation and Resilience Plan from the Spanish State, and by NextGenerationEU from the European Union through the Recovery and Resilience Facility; the La Caixa Banking Foundation, grant no. LCF/BQ/PI21/11830030; the ``Programa Operativo" FEDER 2014-2020, Consejer\'ia de Econom\'ia y Conocimiento de la Junta de Andaluc\'ia (Ref. 1257737), PAIDI 2020 (Ref. P18-FR-1580) and Universidad de Ja\'en; ``Programa Operativo de Crecimiento Inteligente" FEDER 2014-2020 (Ref.~ESFRI-2017-IAC-12), Ministerio de Ciencia e Innovaci\'on, 15\% co-financed by Consejer\'ia de Econom\'ia, Industria, Comercio y Conocimiento del Gobierno de Canarias; the ``CERCA" program and the grant 2021SGR00426, both funded by the Generalitat de Catalunya; and the European Union's ``Horizon 2020" GA:824064 and NextGenerationEU (PRTR-C17.I1).
State Secretariat for Education, Research and Innovation (SERI) and Swiss National Science Foundation (SNSF), Switzerland;
The research leading to these results has received funding from the European Union's Seventh Framework Programme (FP7/2007-2013) under grant agreements No~262053 and No~317446;
This project is receiving funding from the European Union's Horizon 2020 research and innovation programs under agreement No~676134;
ESCAPE - The European Science Cluster of Astronomy \& Particle Physics ESFRI Research Infrastructures has received funding from the European Union’s Horizon 2020 research and innovation programme under Grant Agreement no. 824064.

\clearpage

\section*{Full Author List: CTA-LST Project}
\scriptsize
\noindent
K. Abe$^{1}$,
S. Abe$^{2}$,
A. Aguasca-Cabot$^{3}$,
I. Agudo$^{4}$,
N. Alvarez Crespo$^{5}$,
L. A. Antonelli$^{6}$,
C. Aramo$^{7}$,
A. Arbet-Engels$^{8}$,
C.  Arcaro$^{9}$,
M.  Artero$^{10}$
% ...
\\
$^{1}$Department of Physics, Tokai University.
$^{2}$Institute for Cosmic Ray Research, University of Tokyo.
$^{3}$Departament de Física Quàntica i Astrofísica, Institut de Ciències del Cosmos, Universitat de Barcelona, IEEC-UB.
$^{4}$Instituto de Astrofísica de Andalucía-CSIC.
$^{5}$EMFTEL department and IPARCOS, Universidad Complutense de Madrid.
$^{6}$INAF - Osservatorio Astronomico di Roma.
$^{7}$INFN Sezione di Napoli.
$^{8}$Max-Planck-Institut für Physik.
$^{9}$INFN Sezione di Padova and Università degli Studi di Padova.
$^{10}$Institut de Fisica d'Altes Energies (IFAE), The Barcelona Institute of Science and Technology.
% ...

\section*{Author List: CTA Consortium}
\scriptsize
\noindent
K. Bernlöhr$^{74}$,
O. Gueta$^{75}$,
K. Kosack$^{76}$,
G. Maier$^{75}$,
J. Watson$^{75}$
\\
$^{74}$Max-Planck-Institut für Kernphysik.
$^{75}$Deutsches Elektronen-Synchrotron.
$^{76}$CEA/IRFU/SAp, CEA Saclay.

\end{document}